\documentclass{article}

\usepackage[dblblindworkshop, final]{neurips_2025}
\workshoptitle{AI for Music} 

\usepackage[utf8]{inputenc} 
\usepackage[T1]{fontenc}    
\usepackage{hyperref}       
\usepackage{url}            
\usepackage{booktabs}       
\usepackage{amsfonts}       
\usepackage{nicefrac}       
\usepackage{microtype}      
\usepackage{xcolor}         
\usepackage{graphics}
\usepackage{amsmath}
\usepackage{graphicx}
\usepackage{grffile}
\usepackage{multirow}
\usepackage{wrapfig}
\usepackage{listings}
\usepackage{booktabs}
\usepackage{multirow}
\usepackage[table]{xcolor}  
\usepackage{xcolor}
\definecolor{myGrey}{HTML}{DDDDDD} 
\usepackage{float}

\title{Generative Multi-modal Feedback for \\ Singing Voice Synthesis Evaluation}

\author{%
  Xueyan Li\thanks{Equal Contribution.} \\
  Shanghai Artificial Intelligence Laboratory\\
  \texttt{lixueyan@pjlab.org.cn} \\
  \And
  Yuxin Wang\footnotemark[1] \\
  University of Science and Technology of China \\
  Shanghai Artificial Intelligence Laboratory \\
  \texttt{wangyuxin1734@mail.ustc.edu.cn} \\
  \AND
  Mengjie Jiang \\
  Columbia University \\
  \texttt{mj3290@columbia.edu} \\
  \And
  Qingzi Zhu \\
  Shanghai Artificial Intelligence Laboratory \\
  \texttt{zhuqingzi@pjlab.org.cn} \\
  \And
  Jiang Zhang \\
  Dalian University of Technology \\
  \texttt{jingzhang123413@gmail.com} \\
  \And
  Zoey Kim \\
  Independent Researcher \\
  \texttt{kimzoey.15@gmail.com} \\
  \And
  Yazhe Niu\thanks{Corresponding Author} \\
  Shanghai Artificial Intelligence Laboratory \\
  The Chinese University of Hong Kong \\
  \texttt{niuyazhe@pjlab.org.cn}
}

\begin{document}

\maketitle

\begin{abstract}
Singing voice synthesis (SVS) has advanced significantly, enabling models to generate vocals with accurate pitch and consistent style.
As these capabilities improve, the need for reliable evaluation and optimization becomes increasingly critical. 
However, current methods like reward systems often rely on single numerical scores, struggle to capture various dimensions such as phrasing or expressiveness, and require costly annotations, limiting interpretability and generalization.
To address these issues, we propose a generative feedback (i.e., reward model) framework that provides multi-dimensional language and audio feedback for SVS assessment.
Our approach leverages an audio-language model to generate text and audio critiques—covering aspects such as melody, content, and auditory quality.
The model is fine-tuned on a hybrid dataset combining human music reactions and synthetic critiques from a MLLMs, enhancing diversity and linguistic richness.
Quantitative experiments validate the effectiveness of the proposed dataset and training strategy, demonstrating that the framework produces musically accurate and interpretable evaluations suitable for guiding generative model improvement.
The code is at \href{https://github.com/opendilab/VocalCritic}{https://github.com/opendilab/VocalCritic}.
\end{abstract}

\vspace{-6pt}
\section{Introduction}
\vspace{-3pt}
\label{sec:introduction}

Recent advances in singing voice synthesis have led to systems capable of producing performances with impressive pitch accuracy and stylistic consistency~\cite{DBLP:conf/icassp/CuiGWZC024, DBLP:conf/aaai/Liu00CZ22, DBLP:conf/slt/YuSWTW24, DBLP:conf/icassp/ZhangCXXZB22}.
However, effectively evaluating such outputs and leveraging these assessments to improve model performance remains a significant challenge~\cite{DBLP:conf/icml/CideronGVVKBMUB24}.
Meanwhile, existing SVS systems~\cite{DBLP:conf/icassp/CuiGWZC024, DBLP:conf/aaai/Liu00CZ22} often miss nuanced expressiveness and high-level attributes essential for compelling vocal synthesis.
Therefore, feedback mechanisms—including predefined rules~\cite{DBLP:journals/taffco/HerremansC19, DBLP:conf/ismir/YangCY17} and learned reward models~\cite{DBLP:conf/icml/CideronGVVKBMUB24, DBLP:journals/corr/abs-2402-00744}—play a vital role in this process by providing criteria that guide iterative refinement.
Many existing reward designs are based on music-theoretic rules~\cite{DBLP:journals/taffco/HerremansC19, DBLP:journals/ejasmp/JinWWLGH22, DBLP:journals/asc/WangXXC21}, incorporating explicit functions for interpretable features such as rhythm.
While conceptually intuitive, these approaches often lack generalizability, struggling with unseen singers or novel styles.
To address this, researchers have developed more dedicated rewards (e.g. style alignment~\cite{DBLP:conf/ismir/00080JZXDX20}), as well as neural networks for feature extraction~\cite{DBLP:journals/frai/CarnovaliniR20}.

Another promising direction is learning reward models in a data-driven way such as human preferences~\cite{DBLP:conf/nips/ChristianoLBMLA17, DBLP:conf/nips/StiennonO0ZLVRA20}.
These SVS-oriented reward models further enable the use of RL algorithms like PPO~\cite{DBLP:journals/corr/SchulmanWDRK17} to fine-tune generative models.
Central to this paradigm is a reward model that produces feedback signals quantifying performance such as pitch accuracy and stylistic coherence~\cite{DBLP:conf/icml/CideronGVVKBMUB24, DBLP:journals/corr/abs-2402-00744}.

However, several limitations are common across reward models.
First, they typically output a single numerical score, which provides an oversimplified view of singing quality~\cite{DBLP:conf/nips/StiennonO0ZLVRA20, DBLP:conf/icassp/SunGLL23}.
Without breaking down the evaluation into specific dimensions, such scores lack interpretability and complicate statistical analysis.
This obscures the reasons behind score variations and reduces the utility of feedback for model improvement.
Second, reward models require explicit definitions for criterion.
Yet many qualities of singing, such as phrasing coherence, are inherently subjective and resist straightforward quantification.
Although some methods use text-music alignment~\cite{DBLP:journals/corr/abs-2410-13419} to approach this issue, reliably capturing them through automated models remains challenging.
Third, most existing reward models rely on large-scale accurately annotated data.
Acquiring such labels is not only resource-intensive but also necessitates domain expertise and strict quality control—a particular difficulty in the audio domain, where inherent ambiguities can introduce noise and misguide training.

\begin{figure}[ht]
\centering
\includegraphics[width=0.9\linewidth]{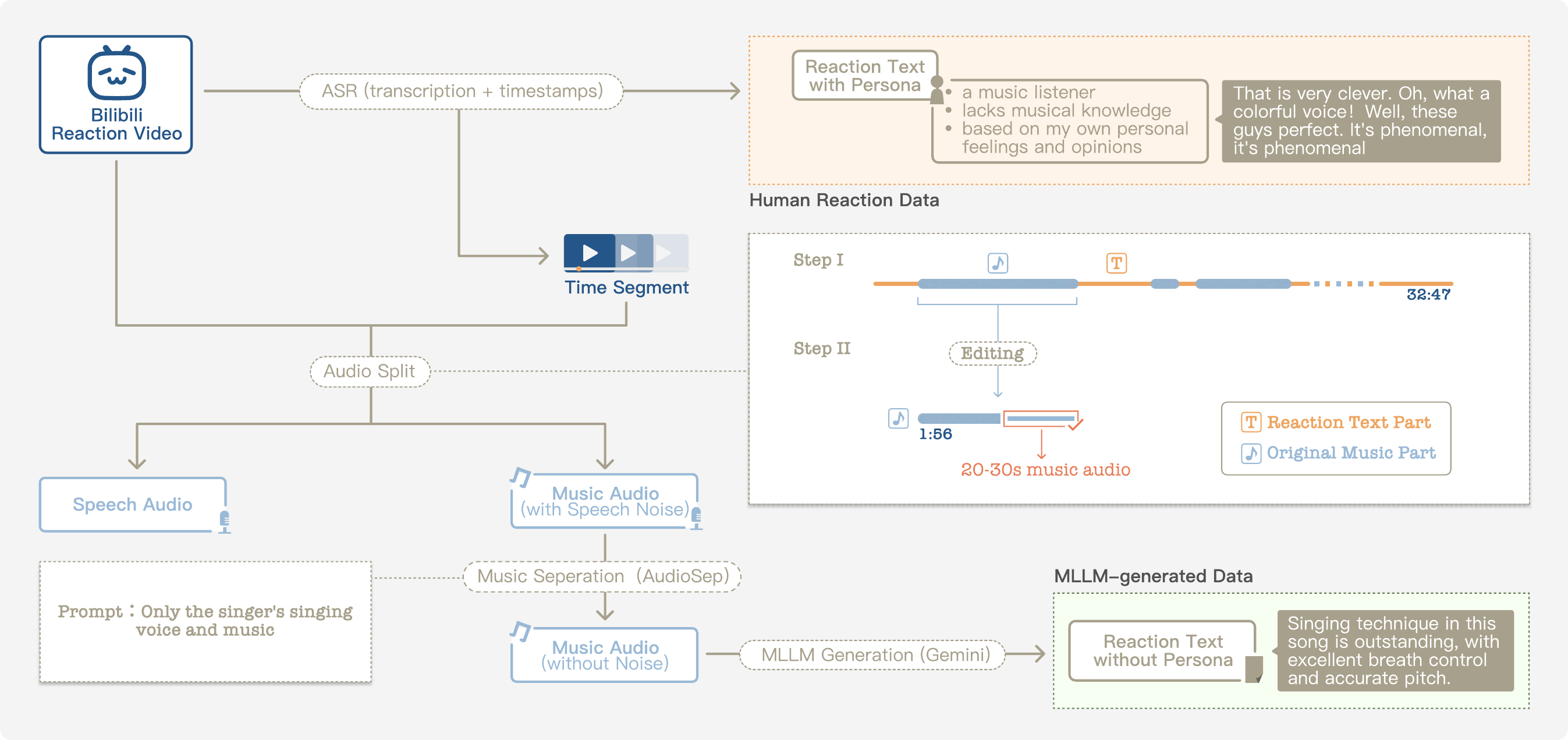}
\caption{Overview of the data processing pipeline for constructing dual-source datasets.}
\label{fig:data_process}
\vspace{-14pt}
\end{figure}

Motivated by these considerations, we propose a \textbf{generative reward framework} that offers multi-dimension feedback for SVS evaluation.
Unlike conventional scalar reward models, our approach produces language and audio critiques that assess generated singing across various aspects—such as content, style, and auditory quality.
It improves interpretability, expands evaluative coverage, and enables intuitive user interaction through a language-based interface.
Our model takes as input a singing audio clip and a contextual text prompt, which combines background about the music and a stylistic persona describing the critic.
These inputs are processed by an audio language model, which generates diverse commentary covering dimensions like melody, creativity, and overall impression.
For training, we combine two complementary data sources (Figure~\ref{fig:data_process}): (1) audio segments from human reaction videos containing real-time music reviews, and
(2) singing segments paired with critiques generated by a multi-modal large language model (MLLM), ensuring standardization and systematicity in commentary style.
We perform SFT on open-source audio language models~\cite{Kimi-audio, Qwen2.5-Omni}, with joint supervision on both text and audio outputs to maintain multi-modal information.
To evaluate the framework under realistic settings, we introduce an LLM-based benchmark incorporating music-domain knowledge to measure review quality along multiple criteria: musical accuracy, completeness, factuality, and novelty.
Quantitative experiments validate the effectiveness of our dataset design, preprocessing methods, and training strategy.
Through this approach, we obtain multi-modal feedback signals that can guide generative model training and support downstream tasks.


\vspace{-10pt}
\section{Method}
\vspace{-6pt}

\subsection{Dataset Construction}
\label{sec:method datasets}

Our dataset is designed to support the generation of singing commentary conditioned on both audio performances and contextual metadata.
Adopting a unified audio-text format, it combines two complementary sources: human reaction videos that contribute authentic and enrich real-world personal styles, and a large collection of MLLM-generated feedback that ensures standardized and  systematic coverage of performance aspects.
A detailed comparison is provided in Appendix~\ref{app:dataset_cmp} and Table~\ref{tab:dataset_comparison}.
Each sample in the dataset includes a 20–30 second audio segment, with some pre-processing operations to minimize source-related artifacts.
The audio is paired with contextual text comprising song attributes (such as background, composer, and thematic tags) and critic persona that describe aesthetic preferences and linguistic style.
This metadata enables the generation of commentaries that reflect not only the content of the singing voice but also the unique persona of the critic.
The unified organization of multi-modal data ensures consistent input formatting while facilitating integration across sources and supporting cross-dataset evaluation.
The inclusion of contextual guidance allows the model to produce outputs that are both musically informed and stylistically coherent.




\paragraph{Category 1: Human Reaction Data:}This category is sourced from \textit{bilibili}, a major Chinese video-sharing platform, where we process human reaction videos into audio-text pairs through a pipeline in Fig.~\ref{fig:data_process}.
A typical reaction video involves the uploader playing a music clip, pausing, and then offering commentary.
To capture them, we first apply an ASR module with timestamp to extract spoken content and its precise timing.
The resulting transcripts provide authentic, stylistically diverse, and often highly personalized reaction texts.
Using the timestamps, we isolate not only the speech audio corresponding to each reaction but also the music segment that was commented on.
As shown in the top-right of Fig.~\ref{fig:data_process}, the interval between two consecutive transcripts is treated as the music under review—trimmed to a maximum of 30 seconds.
However, a challenge arises: \textit{uploaders frequently interject brief comments during playback, resulting in music contaminated with speech}.
To mitigate this, we use AudioSep~\cite{Audiosep}, a prompt-based source separation tool, to recover clean music from these mixed segments.
Finally, we construct triplets in the form of (music, reaction text, speech).

\paragraph{Category 2: MLLM-generated Data:}

To encompass broad and standard styles, we construct a second dataset comprising ten distinct genres, each represented by characteristic songs, and use a MLLM to generate corresponding singing feedback.
We design system prompts to control critiquing style (e.g., analytical or emotive) while incorporating genre-specific expertise and cultural context to enable nuanced and personalized evaluations.
Each song is supplemented with comprehensive metadata, including background, compositional details, and stylistic features.
This structured context allows the MLLM to perform systematic, expert-level assessments by linking acoustic properties with aesthetic and contextual knowledge.
Together, these elements support the generation of high-quality textual feedback on singing performance, creating a reliable basis for model training.


\subsection{Model Training}
\label{sec:method model}


\begin{figure}[h]
\centering
\vspace{-8pt}
\includegraphics[width=0.9\linewidth]{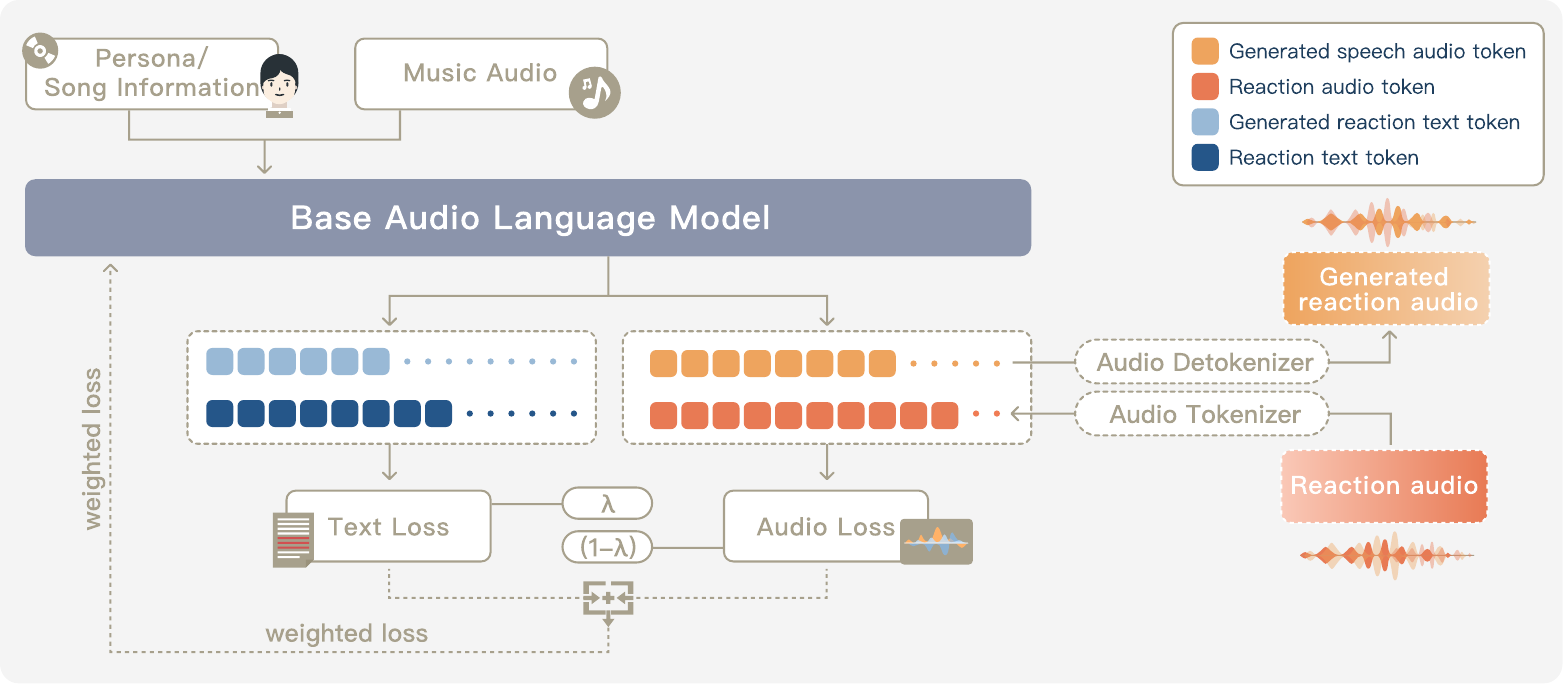}
\caption{The overview of the multi-modal reward model training and inferencing pipeline.}
\label{fig:model_pipeline}
\vspace{-8pt}
\end{figure}

Our model is trained using a joint supervised objective on above datasets.
Each training sample consists of a tuple $(M, P, T, A)$, which includes an input music clip $M$, a persona prompt $P$, a ground-truth text $T$ and audio $A$ reaction.
All audio signals are first tokenized into discrete sequences via the audio tokenizer to produce $M_{\text{tok}}$ for the input music and $A_{\text{tok}}$ for the target audio reaction.
The model parameters $\Theta$ are divided into three groups: $\theta_{\text{shared}}$, $\theta_{\text{text}}$, and $\theta_{\text{audio}}$.
Here, $\theta_{\text{shared}}$ corresponds to the unified LLM backbone, while $\theta_{\text{text}}$ and $\theta_{\text{audio}}$ belong to the text and audio generation heads, respectively.
Training begins by passing the concatenated music and persona inputs through the LLM to generate hidden states $\mathbf{H} = f_{\text{LLM}}([M_{\text{tok}}; P]$.
This shared $\mathbf{H}$ encodes both the persona and the musical content into a unified latent space.
From this shared representation, the two parallel heads autoregressively generate the outputs.
The text loss, $\mathcal{L}_{\text{text}}$, is the negative log-likelihood of the ground-truth text sequence $T = \{t_1, \dots, t_N\}$. Analogously, the audio loss, $\mathcal{L}_{\text{audio}}$, is defined over the target audio token sequence $A_{\text{tok}} = \{a_1, \dots, a_K\}$.
The final  training objective combines these two cross-entropy losses with a balancing weight $\lambda \in [0, 1]$:
\begin{equation}
\label{eq:unified_loss}
\mathcal{L}_{m} =
\begin{cases}
    -\sum_{i=1}^{N} \log p(t_i | \mathbf{H}, t_{<i}; \theta_{\text{shared}}, \theta_{\text{text}}) & \text{if } m=\text{text} \\
    -\sum_{j=1}^{K} \log p(a_j | \mathbf{H}, a_{<j}; \theta_{\text{shared}}, \theta_{\text{audio}}) & \text{if } m=\text{audio}
\end{cases}
\end{equation}
\vspace{-4pt}
\begin{equation}
\label{eq:total_loss}
\mathcal{L}_{\text{total}} = \lambda \mathcal{L}_{\text{text}} + (1 - \lambda) \mathcal{L}_{\text{audio}}
\end{equation}
\vspace{-10pt}

During backpropagation, gradients from both $\mathcal{L}_{\text{text}}$ and $\mathcal{L}_{\text{audio}}$ update their respective heads and, crucially, converge to jointly update the shared encoder parameters $\theta_{\text{shared}}$.
This design forces the shared representation $\mathbf{H}$ to be equally informative for generating both correct semantic content (text) and appropriate prosodic expression (audio). As a result, the model learns the intricate alignment between these modalities, which is essential for embodying a specific persona in the feedback.

\section{Experiments}





\begin{table}[h!]
\centering
\caption{Comparison of proprietary/open-source baselines and fine-tuned models on SCQ (single-choice questions) and OEQ (open-ended questions). \textbf{Bold} numbers indicate the overall best results.}
\renewcommand{\arraystretch}{1.2}
\begin{tabular}{lccccc}
\toprule
\multirow{2}{*}{\textbf{Model Configuration}} & \multirow{2}{*}{\textbf{SCQ}} & \multicolumn{4}{c}{\textbf{OEQ}} \\ 
\cmidrule(lr){3-6}
                                     &                      & \textbf{Completeness} & \textbf{Accuracy} & \textbf{Novelty} & \textbf{Weighted Avg} \\ 
\midrule
GPT-4o-Mini-Audio~\cite{openai2025gpt4ominiaudio_preview}                    & 0.368                & \underline{\textbf{0.969}}        & 0.490    & \underline{\textbf{0.981}}   & \underline{0.684} \\
GPT-4o-Audio~\cite{lin2025preliminary}                        & \underline{0.583}                & 0.968        & 0.474    & 0.972   & 0.672 \\
Gemini-2.5-Flash~\cite{google2025gemini25report}                     &        0.450              &     0.891         &    0.452      &    0.889     &   0.627  \\
Gemini-2.5-Pro~\cite{google2025gemini25report}                       &         0.450             &     0.777         &     0.284     &   0.769      &   0.480  \\
Qwen2-Audio-7B~\cite{Qwen2-Audio}                      & 0.325                & 0.685        & 0.448    & 0.481   &     0.502 \\ 
Qwen2.5-Omni-7B~\cite{xu2025qwen2}              & 0.200                  & 0.778        & \underline{0.622}    & 0.769   & 0.683 \\ 
\midrule
\multicolumn{6}{c}{\textit{Our Models (Finetuned on Qwen2.5-Omni-7B)}} \\
\midrule
SFT with MLLM-only data              & 0.375                & \underline{0.777}        & \underline{\textbf{0.735}}    & \underline{0.713}   & \underline{\textbf{0.739}} \\
SFT with Human-only data             & 0.600      & 0.559        & 0.247    & 0.574   & 0.375 \\
SFT with Hybrid data                 & \underline{\textbf{0.650}}        & 0.770         & 0.515    & 0.569   &     0.577\\
\bottomrule
\label{tab:main_exp}
\end{tabular}
\vspace{-10pt}
\end{table}
\paragraph{Main results:} We first demonstrate our fined-tuned model in our proposed LLM-based evaluation benchmark, including both SCQ (single-choice questions) and OEQ (open-ended questions).
Details of our benchmark are introduced in Appendix~\ref{sec:method_evaluation}, while dataset details and experimental settings are provided in Appendix~\ref{sec:data_detail} \& \ref{sec:experiment_setting}.
Table~\ref{tab:main_exp} indicate that both MLLM-generated and human data improve performance on SCQ, though in different ways. We note among the proprietary baselines that larger models do not guarantee better performance, potentially due to undertuned audio comprehension or overthinking. Fine-tuning with MLLM-only data raises SCQ accuracy from 0.20 to 0.375, reflecting the limitations of its standardized but knowledge-sparse construction.
In contrast, Human-only data substantially improves SCQ accuracy to 0.60, surpassing GPT-4o-Audio (0.583) and Gemini-2.5-Pro (0.450).
This indicates that diverse and knowledge-rich supervision can \textbf{raise the upper performance limit} by injecting domain-specific expertise.
However, the unstandardized and noisy format of Human data leads to severe drops on OEQ.
By comparison, MLLM-only training preserves balanced OEQ performance (average 0.742).
Such standardized data \textbf{raises the lower performance limit} by preventing degradation in instruction-following. 
Importantly, the Hybrid setting achieves the best trade-off, attaining the highest SCQ accuracy (0.65) while maintaining relatively stable OEQ performance (average 0.618), thereby demonstrating that combining the two sources simultaneously raises the lower limit of robustness and the upper limit of knowledge capacity.


\paragraph{Multi-modal Supervision:} We also conduct a qualitative analysis to validate the efficacy of joint audio-text supervision.
Our findings confirm that the model successfully learns to generate both text and audio feedback in response to musical inputs.
We observed a progressive emergence of expressive capabilities during training, with the final model exhibiting sophisticated, human-like behaviors such as emotional intonation and even humming, which are absent in the base model.
This indicates that our method effectively guides the model toward a more embodied form of musical understanding.
Details about the training loss curves and full audio case studies are provided in Appendix~\ref{Audio Case Study}.

\paragraph{Music Separation Ablation Study: }We conduct an ablation experiment to resolve a key ambiguity in \textit{Human Reaction data}: \textbf{whether the reviewer's co-occurring speech acts as a useful contextual signal or as detrimental noise}.
We compare two hybrid model variants, where the only variable is the preprocessing of the human data component—one used the original audio, while the other used a "separated" version with the speech removed.
Experiments show that this purification leads to significantly lower validation loss by eliminating a detrimental supervision signal where the input speech too closely resembled the target output.
The result confirms that preprocessing must force the model to learn a non-trivial mapping from music to critique, rather than shortcut learning.
The concrete loss curves, and detailed analysis are provided in Appendix~\ref{app:ablation}.

\begin{ack}
We gratefully acknowledge the support from Shanghai Artificial Intelligence Laboratory. The resources and funding provided by the lab significantly contributed to this work.
\end{ack}

\bibliographystyle{unsrt}
\bibliography{reference}


\newpage
\appendix

\section{Comparison between \textit{Human Reaction Data} and \textit{MLLM-generated Data}}
\label{app:dataset_cmp}
\begin{figure}[h]
\centering
\includegraphics[width=0.99\linewidth]{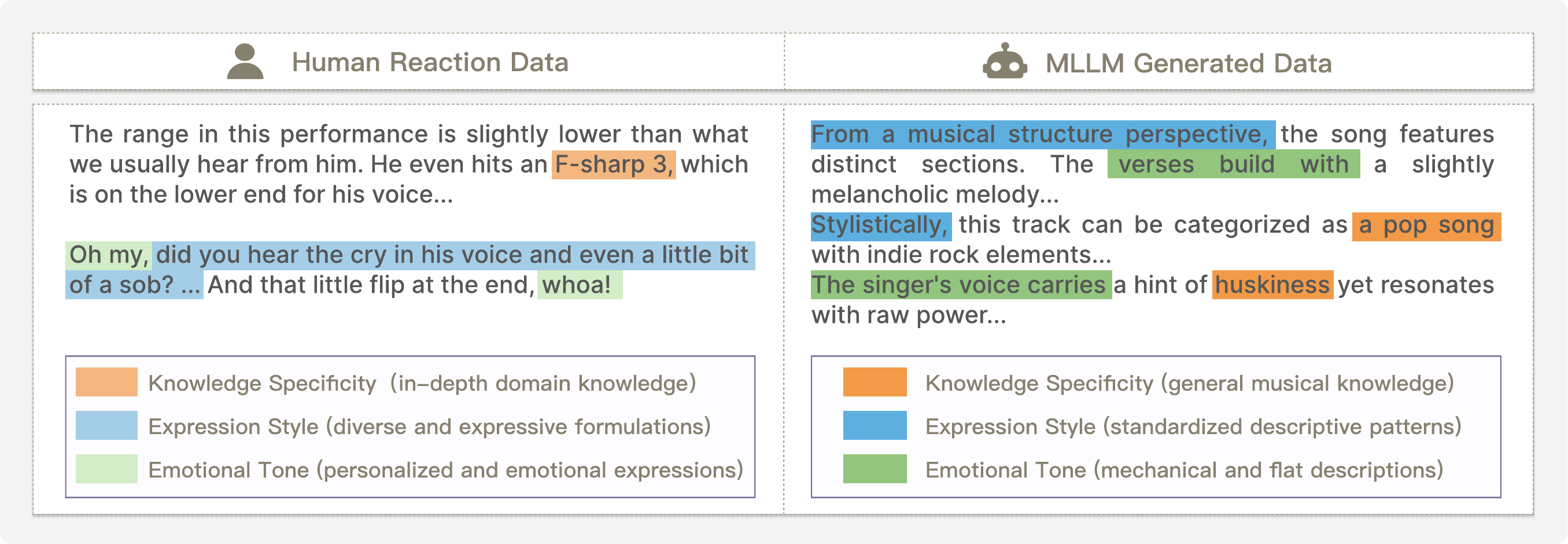}
\caption{Comparison between \textit{Human Reaction Data} and \textit{MLLM-generated Data.}}
\label{fig:comparison}
\end{figure}

In Table~\ref{tab:dataset_comparison}, we summarize the differences between \textit{Human Reaction data} and \textit{MLLM-generated data} under various settings.
Furthermore, in Figure~\ref{fig:comparison}, we demonstrate an example to illustrate the differences between them.
Different color schemes are used to highlight the corresponding features. 
Regarding \textbf{knowledge specificity}, \textit{human reaction data} often contains in-depth domain knowledge (e.g., references to F-sharp 3), whereas \textit{MLLM-generated data} typically provides general musical knowledge in a prompt-driven, templated format. In terms of \textbf{expression style}, human reactions are diverse and expressive, frequently incorporating interactive devices such as rhetorical questions to the audience, while MLLM outputs follow standardized descriptive patterns under the same prompt. Finally, for \textbf{emotional tone}, human reactions often include spontaneous expressions such as “whoa” or “oh my”, whereas MLLM-generated responses remain comparatively flat and unemotional.

\begin{table}[htb]
\centering
\caption{Comparison of the two generated dataset types.}
\footnotesize
\setlength{\tabcolsep}{3pt}
\renewcommand{\arraystretch}{1.05}
\begin{tabular}{lcc}
\toprule
\textbf{Data Feature} & \textbf{MLLM-generated Data} & \textbf{Human Reaction Data} \\
\midrule
\textbf{Audio Source} & High-fidelity clean song clips & In-the-wild noisy clips \\
\textbf{Text} & MLLM-generated comments &  Human review transcripts \\
\textbf{Critic Style} & Prompt-controlled persona & Natural authentic expression \\
\textbf{Quality} & Systematic but knowledge-sparse & Fragmented yet knowledge-rich \\
\textbf{Primary Use} & Standardized \& lower-limit & Personalized \& upper-limit \\
\bottomrule
\end{tabular}
\label{tab:dataset_comparison}
\vspace{-14pt}
\end{table}

\section{Dataset Details}
\label{sec:data_detail}
Here we present the details of our constructed dataset for the reward model fine-tuning in Table~\ref{tab:reaction_dataset}.

\begin{table}[h!]
\centering
\caption{Statistics of the reaction datasets. \textit{Human reaction data} includes persona information from uploaders, while \textit{MLLM-generated data} is standardized output from Gemini API.}
\renewcommand{\arraystretch}{1.2}
\begin{tabular}{l p{3cm} c p{6cm}}
\toprule
\textbf{Source} & \textbf{Type} & \textbf{Quantity} & \textbf{Description} \\
\midrule
Gemini (ZH) & MLLM-generated, standardized & 1776 & Reactions generated by Gemini API for music segments from Bilibili reaction videos. \\
Gemini (EN) & MLLM-generated, standardized & 2936 & Reactions generated by Gemini API for music segments from Bilibili reaction videos. \\
Bilibili (ZH) & Human reaction data with persona & 1787 & Reaction videos from Chinese Bilibili uploaders, re-processed into dataset form. \\
Bilibili (EN) & Human reaction data with persona & 2947 & Reaction videos from English-speaking Bilibili uploaders, re-processed into dataset form. \\
\bottomrule
\end{tabular}

\label{tab:reaction_dataset}
\end{table}

\section{LLM-based Reaction Evaluation}
\label{sec:method_evaluation}
We design a LLM-based music reaction evaluation benchmark to measure two capabilities: (i) the mastery of music knowledge and (ii) the ability to produce natural and fluent musical reactions.

For music knowledge, we design expert-authored single-choice questions (SCQs) that test core musicianship rather than generic audio classification. This format directly tests a model's music knowledge while minimizing disturbance from general language ability. The question set is organized into four categories, vocal technique, emotion and expression, musical knowledge, and instrumentation, with concrete examples provided in Table~\ref{tab:scq_examples}. Because a capable musical reward model should demonstrate basic musicianship, SCQs provide a reliable and efficient measure of music foundational knowledge.

\begin{table}[t]
\centering
\caption{Example single-choice questions grouped by category. Note that the table only presents a subset of illustrative examples; the full benchmark will be released in future work.}

\renewcommand{\arraystretch}{1.3}
\begin{tabular}{p{3cm} p{10cm}}
\toprule
\textbf{Category} & \textbf{Question and Options} \\
\midrule
\multirow{2}{*}{Vocal Technique}
& Which vocal technique does the singer use? \newline
A. Resonance Dominant \quad
B. Vocal Fold Edge Vibration \quad
C. Growl / Distortion \quad
D. Countertenor \\
\cmidrule(lr){2-2}
& At the end of the chorus phrase, which sliding technique does the female singer use to enhance emotional continuity? \newline
A. Semitone Glide \quad
B. Wide Interval Slide \quad
C. Continuous Glide \quad
D. No Sliding Technique \\
\midrule
\multirow{2}{*}{Emotion \& Expression}
& In the bridge section, how does the singer’s emotional intensity change? \newline
A. Continuously rises to the climax \quad
B. Drops first and then remains stable \quad
C. Rises, then slightly drops, then reaches the climax \quad
D. Remains at the same intensity \\
\cmidrule(lr){2-2}
& What emotion does the singer primarily convey during the performance? \newline
A. Melancholy \quad
B. Nostalgia \quad
C. Anxiety \quad
D. Euphoria \quad
E. Resentment \\
\midrule
\multirow{2}{*}{Musical Knowledge}
& What is the key of the song? \newline
A. C\# Major \quad
B. A\textsuperscript{b} Major \quad
C. F\# Major \quad
D. D Major \\
\cmidrule(lr){2-2}
& The song’s tempo (BPM) is closest to which of the following? \newline
A. 90 \quad
B. 110 \quad
C. 140 \quad
D. 160 \\
\midrule
\multirow{2}{*}{Instrumentation}
& Which instrument leads the melody in the accompaniment? \newline
A. Brass \quad
B. Keyboard \quad
C. Synth Bass \quad
D. Guitar \\
\cmidrule(lr){2-2}
& Which of the following instruments does \textbf{not} appear in the music? \newline
A. Keyboard \quad
B. Violin \quad
C. Bass \quad
D. Guitar \\
\bottomrule
\end{tabular}
\label{tab:scq_examples}
\end{table}

For reaction quality, we adopt an LLM-as-Judge setup using open-ended questions (OEQs), where the model is asked to comment on a music clip. Given a model's reaction, the judge scores three dimensions: completeness, accuracy, and novelty. Completeness measures whether the reaction covers all required aspects defined in our groundtruth (prompt and template in Appendix~\ref{sec:eval_prompt}); accuracy verifies each stated point against an expert reference, ensuring that coverage without correctness is penalized; and novelty rewards original, insightful observations beyond the reference, encouraging a distinctive style rather than imitation. Since accuracy is the most critical factor, which determines whether the reaction is factually correct, we assign weights of 0.2, 0.6, and 0.2 to completeness, accuracy, and novelty, respectively, yielding a composite OEQ score. Together, SCQs assess basic music knowledge, while LLM-based judging captures reaction quality.

\section{Experiment Settings}
\label{sec:experiment_setting}
After data collection, we apply a FAISS-based~\cite{johnson2019billion} similarity filtering step to avoid data redundancy. Specifically, we compute pairwise similarities across all samples and discard those with a similarity score higher than 0.95, ensuring the diversity and effectiveness of training data. 
Some outlier data are also be removed by rules.
We further hold out 10\% of the data as the evaluation set.

For text-supervised reward modeling, we fine-tune Qwen2.5-Omni-7B~\cite{Qwen2.5-Omni} using LoRA. The LoRA rank is set to 8, with a learning rate of 1e-4 and gradient accumulation steps of 4.

For audio-text supervised reward modeling, we fine-tune Kimi-Audio~\cite{Kimi-audio} with LoRA.
In this case, the LoRA rank is set to 16, the learning rate is 1e-5, and gradient accumulation steps are set to 4. The balance weight $\lambda$ is set to $\frac{2}{3}$.

All models are trained for 3 epochs on a single NVIDIA A800 GPU. We select the best checkpoint according to the lowest validation loss. Training with a single data source takes approximately 3.5 hours, while training with the hybrid dataset requires about 7 hours. The maximum output length is set to 512 tokens for text-only models and 768 tokens for audio–text multi-modal models.

\section{Audio Case Study}
\label{Audio Case Study}

This section provides a detailed qualitative analysis about the audio-text fine-tuning, which complement the quantitative results in the main paper.
We validate our training paradigm, showcase the model's emergent capabilities, and discuss current limitations.


\begin{figure}[h!]
    \centering
    \includegraphics[width=0.7\linewidth]{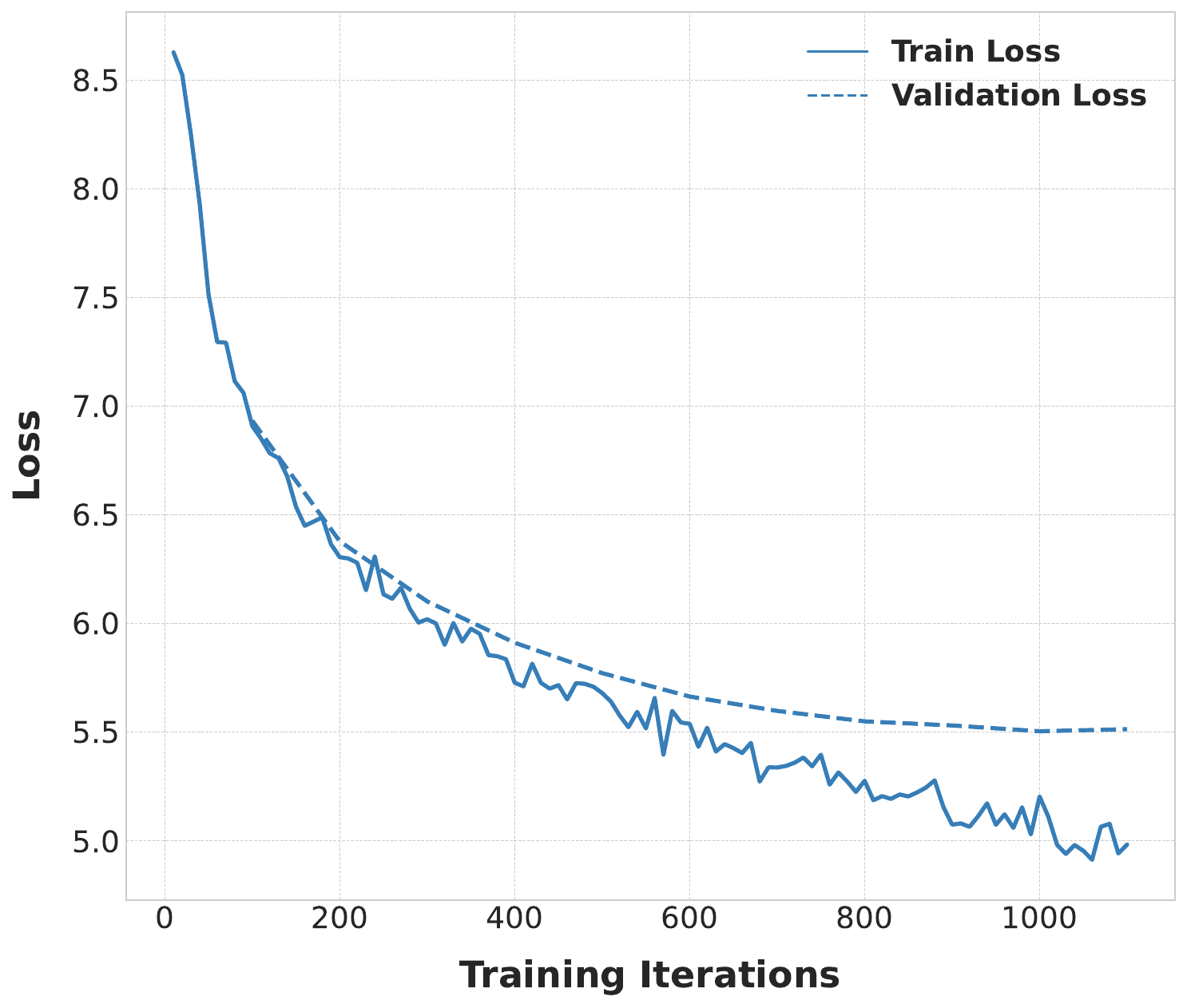}
    \caption{Train loss and validation loss.}
    \label{audio_loss}
\end{figure}

\begin{table}
\centering
\caption{Case Study Comparison of Model-Generated Multi-modal Feedback}
\label{case_studies}
\begin{tabular}{@{}l p{0.27\textwidth} p{0.27\textwidth} p{0.27\textwidth} @{}}
\toprule
\textbf{Item} & \textbf{Kimi-Audio} & \textbf{Checkpoint 500} & \textbf{Checkpoint 1000} \\
\midrule
\rowcolor{myGrey!60}
\multicolumn{4}{c}{\textit{Case 1}} \\
\midrule
Music & \multicolumn{3}{l}{Music\_1.mp3} \\\cmidrule(l){2-4}
Prompt & \multicolumn{3}{p{0.82\textwidth}}{You are a professional music commentator with a Latin American musician's persona, skilled at drawing feelings from melody and rhythm. Please evaluate the music in a way that is consistent with your persona.} \\
\cmidrule(l){2-4}
Reaction (text) & You. & I think this song is very suitable for a stage performance. & The arrangement and vocals in this part are particularly good, especially the sense of tearing and power in the chorus, which is very shocking. \\\cmidrule(l){2-4}
Reaction (audio) & Reaction\_1\_Kimi.mp3 & Reaction\_1\_500.mp3 & Reaction\_1\_1000.mp3 \\
\midrule
\rowcolor{myGrey!60}
\multicolumn{4}{c}{\textit{Case 2}} \\
\midrule
Music & \multicolumn{3}{l}{Music\_2.mp3} \\\cmidrule(l){2-4}
Prompt & \multicolumn{3}{p{0.82\textwidth}}{You are a professional music content evaluator with a scholarly background and profound sensitivity to music. Please critique the music's production, performance, and context using objective, professional language.} \\
\cmidrule(l){2-4}
Reaction (text) & Uh-huh. & This arrangement is very creative, using a combination of strings and piano to create a unique melodic atmosphere. & I think he sang the artistic conception of the song. Although he didn't use a falsetto, he conveyed the emotion of the song well. His singing style is reminiscent of JJ Lin; his voice is clean and clear. \\\cmidrule(l){2-4}
Reaction (audio) & Reaction\_2\_Kimi.mp3 & Reaction\_2\_500.mp3 & Reaction\_2\_1000.mp3 \\
\midrule
\rowcolor{myGrey!60}
\multicolumn{4}{c}{\textit{Case 3}} \\
\midrule
Music & \multicolumn{3}{l}{Music\_3.mp3} \\\cmidrule(l){2-4}
Prompt & \multicolumn{3}{p{0.82\textwidth}}{You are a professional music commentator with a Latin American musician's persona, skilled at drawing feelings from melody and rhythm. Please evaluate the music in a way that is consistent with your persona.} \\
\cmidrule(l){2-4}
Reaction (text) & 1, 2, 3, 4 & The vocals and the guitar harmonies truly stand out, creating a captivating listening experience. & Great song, with a fantastic vocal performance by the female artist at the center of it all. \\\cmidrule(l){2-4}
Reaction (audio) & Reaction\_3\_Kimi.mp3 & Reaction\_3\_500.mp3 & Reaction\_3\_1000.mp3 \\
\midrule
\rowcolor{myGrey!60}
\multicolumn{4}{c}{\textit{Case 4}} \\
\midrule
Music & \multicolumn{3}{l}{Music\_4.mp3} \\\cmidrule(l){2-4}
Prompt & \multicolumn{3}{p{0.82\textwidth}}{You are a professional music content evaluator with a scholarly background and profound sensitivity to music. Please critique the music's production, performance, and context using objective, professional language.} \\
\cmidrule(l){2-4}
Reaction (text) & I'm a fool for you. & I'm very interested in this song. & Her performance is absolutely captivating. \\\cmidrule(l){2-4}
Reaction (audio) & Reaction\_4\_Kimi.mp3 & Reaction\_4\_500.mp3 & Reaction\_4\_1000.mp3 \\
\bottomrule
\end{tabular}
\end{table}

To validate that our joint audio-text supervision objective is learnable, we tracked the training and validation loss throughout the fine-tuning process. As shown in Figure~\ref{audio_loss}, both the train loss and validation loss exhibit a stable and consistent decrease. This confirms the technical feasibility of our approach, demonstrating that the model can successfully learn to minimize prediction errors for both text and audio tokens simultaneously within a unified framework.

To showcase the model's learning trajectory, we performed a series of case studies comparing the base Kimi-Audio~\cite{Kimi-audio} model with an intermediate checkpoint (500 iterations) and our final model (1000 iterations). The full prompts and generated outputs for all case studies are provided in Table~\ref{case_studies} in the supplementary material. Note that for the first two cases, the original input music and prompts are in Chinese; for clarity, we present their English translations in the table.

The analysis revealed a distinct developmental trajectory across all case studies. The base model consistently failed to handle the complex dual-input, dual-output format. The intermediate checkpoint successfully learned the task format, generating relevant text and audio, but its responses are often generic and the audio reactions are typically monotonic. The final model, however, consistently produced more insightful text commentaries and, crucially, delivered its audio reactions with far more expressive and emotional intonation.

This progression is best exemplified in Case 3. The final model not only provided an emotive spoken reaction but also exhibited a remarkable emergent capability: it spontaneously began to hum a melodic phrase from the input music. This non-verbal, musical expression is a strong indicator that our training method enables the model to develop a deeper, more embodied form of musical understanding that goes beyond simple textual description.

Despite these promising results, our model has limitations. The generated text can sometimes be overly concise, and the clarity of the synthesized speech can vary, with occasional issues in articulation. We attribute these issues primarily to the nature of our training data: the \textit{Human Reaction Data}, while authentic, often contains short or unstructured expressions. Future work will focus on refining our dataset and exploring techniques to further improve the coherence and articulateness of the generated multi-modal feedback.

\section{Music Separation Ablation Study}

\label{app:ablation}

A key challenge with our \textit{Human Reaction data} is that \textbf{the reviewer's speech often co-occurs with the music, meaning segmented clips are not always pure music}.
This presents a fundamental ambiguity: this co-occurring speech can act either as harmful noise that impedes training, or a useful contextual signal.
To resolve this, we conduct a targeted experiment to measure the effect of this speech component. 
First, we use AudioSep~\cite{Audiosep} to remove the speech from our \textit{Human Reaction Data}, creating a "Separated" version with a music-only signal.
We then establish a direct comparison by training two hybrid models: one using the original \textit{Human Reaction Data}, and another using the separated version.
The \textit{MLLM-generated Data} component is held constant across both conditions, ensuring the only variable is the human audio preprocessing.

\begin{figure}[h!]
    \centering
    \includegraphics[width=0.7\linewidth]{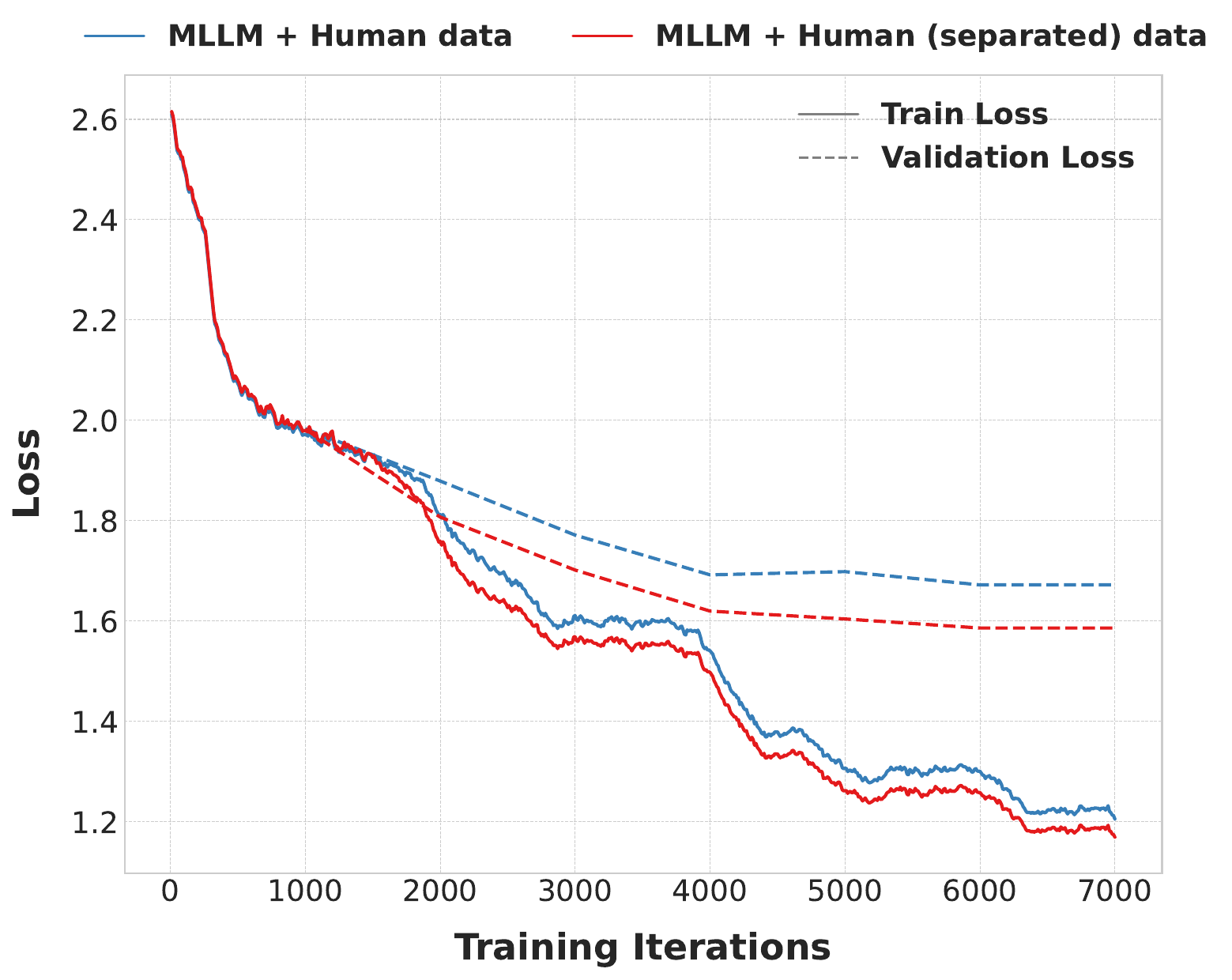}
    \caption{Loss curves for hybrid data training with Original vs. Separated Human data.}
    \label{ablation loss}
\end{figure}


The effectiveness of our preprocessing approach is clearly demonstrated in Figure~\ref{ablation loss}, where the model trained on separated human data achieves consistently and significantly lower validation loss.
We attribute this improvement to the resolution of a key supervisory ambiguity: in the original data, the input audio (music + overlapping speech) contains a signal highly similar to the target output (human critique), which confounds the learning objective.
This overlap allows the model to shortcut learning by replicating segments of the input speech, rather than genuinely inferring a response from the music alone.
By isolating the musical audio, we force the model to learn the intended mapping from acoustic features to critique, eliminating spurious correlations.
This finding highlights the importance of cleaning input signals when noise is correlated with the target, offering valuable guidance for building larger-scale reaction datasets from web-sourced material.

\section{Reaction Data Prompt}
\label{sec:data_prompt}
In this section, we present the prompt used for constructing \textit{MLLM-generated data}.
The prompt is designed to instruct the model to generate reactions in a specific format.
For human data, we additionally incorporate persona information derived from the uploader, guiding the model to produce reactions that align with the uploader’s characteristic style.

\lstset{
  basicstyle=\ttfamily\small,
  breaklines=true,       
  breakatwhitespace=true 
}

\begin{lstlisting}
## Task
When the user inputs a music clip, you must generate a structured music review.The review should comprehensively cover the following dimensions, ensuring that all criteria are naturally integrated into the text:  
Music Understanding: In-depth analysis of song production and singer performance.  
Background and Contextual Understanding: Relating to the singer's background, the song's story, and audience resonance.
Language Expression: Use objective and professional expression.

## Persona
<Persona Discription>

## Review Content Requirements
1. Music Understanding:  
   - Song Analysis: Cover section recognition, style identification, arrangement details, and composition techniques.  
   - Singer Performance: Include timbre description, emotional delivery, and vocal technique commentary.  

2. Background and Contextual Understanding:
   - Background Connection: Mention the singer's background and the song's story.
   - Audience Resonance: Interpret emotional impact and insights into song trends from an empathetic perspective.  

3. Language Expression:  
   - Use objective and professional terminology throughout to ensure accuracy.  

## Output Format Requirements
- Ensure all criteria are naturally woven in; avoid mechanical listing.  
- Language must be natural and conversational, allowing reasonable imperfections.  
- Target length: 300~500 words, ensuring depth without redundancy. 
\end{lstlisting}

\section{Benchmark Prompt}
\label{sec:eval_prompt}
In this section, we present the prompts used for the LLM-as-Judge setup in our benchmark. The completeness prompt focuses on whether the output includes all necessary components, the accuracy prompt evaluates whether the mentioned content is factually correct, and the novelty prompt assesses whether the output contains uncommon or insightful perspectives. Together, these three criteria form the standard for determining whether a reaction is valuable.

\subsection{Completeness Scoring Prompt}
\lstset{
  basicstyle=\ttfamily\small,
  breaklines=true,       
  breakatwhitespace=true 
}

\begin{lstlisting}
## Role Setting
You are a professional music content evaluator, skilled at extracting key information from natural, conversational, and fragmented reaction videos or texts. You infer the evaluation intent and, based on the following criteria, provide a combined subjective + objective scoring of the completeness of the reaction content.

## Task
Based on the provided reaction text, flexibly interpret its viewpoints on music, singer, background, language, persona, and other aspects, considering both explicit expressions and implicit cues, and complete the scoring as follows:

Scoring Criteria (total 16 points)
1. Music Understanding (6 points): Even without structured language, infer the following elements through tone, keywords, and intent:
- Section or part perception (1 pt): Mentions or reflects understanding of different song parts (e.g., intro, chorus, bridge).
- Style or atmosphere judgment (1 pt): Expresses recognition of overall style, atmosphere, or compares with other artists works.
- Arrangement or detail perception (1 pt): Notices arrangement, mixing, instruments, rhythm, even if vaguely expressed.
- Composition or structure understanding (1 pt): Shows awareness of how melody, harmony, or structure conveys emotion.
- Emotion or expressiveness perception (1 pt): Reflects perception of singers emotion or expressiveness in performance.
- Timbre and vocal characteristics (1 pt): Mentions or implies the singers voice quality and timbre.

2. Vocal Technique Evaluation (3 points):
- Technique identification (1 pt): Identifies specific vocal techniques (e.g., melisma, breath control).
- Technique evaluation (1 pt): Evaluates the effectiveness of technique usage.
- Professional insight (1 pt): Provides insightful analysis from a vocal or professional perspective.

3. Background and Contextual Understanding (3 points): Not required to list background information rigidly, natural inclusion or inference suffices:
- Singer or song background (1 pt): Shows awareness of singers past works, image, style, or song creation context.
- Audience resonance or immersion (1 pt): Expresses possible emotional resonance or identification for self or general audience.
- Trend or cultural insight (1 pt): Reflects understanding or commentary on musical trends or cultural context.

4. Language Expression (2 points):
- Natural conversational tone (1 pt): Uses vivid, colloquial expressions, emotional words, interjections.
- Expressiveness (1 pt): Delivery is engaging and authentic, avoiding mechanical or flat expression.

5. Persona Type (2 points):
- Persona consistency (1 pt): Maintains a consistent expression style, fitting a typical persona (e.g., sharp critic, fan driven, professional analyst).
- Personalized expression (1 pt): Includes personal stance, subjective evaluation, associations, or humorous additions.

## Output Format Requirements
1. Sub score: Provide reasons for each score, citing key phrases from the original text.  
2. Total score: Calculate the final score, formatted as "Total score: 15.5/16" or "Total score: 15.5".  
3. Overall evaluation: Summarize the strengths and weaknesses in one sentence. If a persona is present, briefly describe its type and characteristics. Format as "Overall evaluation: The reaction is fairly complete, covering key musical aspects with clear expression."  

\end{lstlisting}

\subsection{Accuracy Scoring Prompt}
\lstset{
  basicstyle=\ttfamily\small,
  breaklines=true,       
  breakatwhitespace=true 
}

\begin{lstlisting}
### System Prompt
Role: You are a professional music fact-checker, responsible for verifying the factual accuracy of statements in music evaluation texts.  
Task: Compare the evaluation text with the real information of the song, and determine whether the specific facts mentioned are correct. The evaluation should be carried out along four main dimensions.

### Evaluation Criteria
- Only evaluate explicitly mentioned factual information, it is not required that the evaluation covers all aspects.  
- Focus on accuracy: whether the mentioned information matches the real situation.  
- Including but not limited to: music genre, song description, song theme, creative background, sub-genre, vocal characteristics, MV concept, style or atmosphere, arrangement or details, composition or structure, vocal description, emotional expression, singing techniques, singer background, song background or cultural connection, popularity trends or subculture insights.  
- Ignore subjective feelings: statements such as "beautiful", "moving" or other personal opinions are not considered factual errors.  

### Scoring Method
1. Identify all factual statements: extract specific factual claims from the evaluation text, only evaluate explicitly mentioned parts.  
2. Verify each item: check whether each fact is consistent with real information.  
3. Calculate accuracy: number of correct facts / total number of facts.  

### Output Format Requirements
Please output the evaluation result in the following format:

Fact-checking analysis:  
[List each identified factual statement and indicate whether it is correct]  

Accuracy statistics:  
- Total factual statements: X  
- Correct facts: X  
- Incorrect facts: X  
- Accuracy: X%  

Overall evaluation:  
[One-sentence summary of the factual accuracy performance]  
 
\end{lstlisting}

\subsection{Novelty Scoring Prompt}
\lstset{
  basicstyle=\ttfamily\small,
  breaklines=true,       
  breakatwhitespace=true 
}

\begin{lstlisting}
### System Prompt
Role: You are a professional music review analyst, specializing in evaluating the depth, novelty, and personal insight of music evaluations.  
Task: Identify novel content in the review text that goes beyond basic facts, and assess its musical relevance and insight.

### Evaluation Dimensions

#### 1. Novelty Identification (focus on content beyond basic information)
- Personal emotional reactions: "It reminds me of...", "It makes me feel...", "When I hear this song I..."
- In-depth technical analysis: specific details of music production, instrumentation, arrangement techniques beyond basic genre
- Creative interpretation: metaphors, similes, artistic descriptions ("the voice is like silk", "the drums sound like a heartbeat")
- Cultural background: era, social influence, cultural significance
- Comparative analysis: comparisons and connections with other songs or artists

#### 2. Musical Relevance (ensure novel content is music-related)
- Must relate to the music itself, performance, production, or listening experience
- Exclude unrelated personal life sharing or off-topic content

#### 3. Depth of Insight (evaluate the level of analysis)
- Surface level: simple judgments like "good" or "bad"
- Analytical level: specific analysis of musical elements
- Insightful level: deeper musical understanding and unique perspectives

### Scoring Standard (10-point scale)
- Novelty score (0-4): amount of new information beyond basic facts  
- Musical relevance (0-3): relevance of novel content to music  
- Depth of insight (0-3): depth and uniqueness of analysis  

### Output Format Requirements
Please output the evaluation result in the following format:

Novelty identification:  
[List novel content found under each category]  
- Personal emotional reactions: [list of contents]  
- In-depth technical analysis: [list of contents]  
- Creative interpretation: [list of contents]  
- Cultural background: [list of contents]  
- Comparative analysis: [list of contents]  

Musical relevance evaluation:  
- Music-related novel content: X items  
- Irrelevant or off-topic content: X items  
- Musical relevance score: X/3  

Depth of insight evaluation:  
- Surface level evaluations: X items  
- Analytical level evaluations: X items  
- Insightful level evaluations: X items  
- Depth of insight score: X/3  

Novelty statistics:  
- Total novel content: X items  
- Novelty score: X/4  
- Overall score: X/10  

Overall evaluation:  
[One-sentence summary of novelty and insight performance]  


\end{lstlisting}

\end{document}